\documentstyle[11pt]{article}
\begin{document}
\centerline{\Large\bf Quantum bit commitment in a noisy channel}
\vskip 2mm
\centerline{S.N.Molotkov and S.S.Nazin}
\centerline{\sl\small Institute of Solid State Physics 
of Russian Academy of Sciences}
\centerline{\sl\small Chernogolovka, Moscow District, 142432, Russia}
\vskip 3mm

\begin{abstract}
Under rather general assumptions about the properties of a noisy quantum 
channel, a first quantum protocol is proposed which allows to implement the 
secret bit commitment with the probability arbitrarily close to unity.
\end{abstract}
\vskip 1mm

The idea that quantum physics can provide more secure communication
between two distant parties than the classical one was first put forward
by Wiesner [1]. Later, after the works [2,3], a lot of papers devoted
to secret key distribution (quantum cryptography) have been published.
Apart from the key distribution protocol, there exist other cryptographic
protocols which are both important for applications and interesting 
in themselves. These are the so-called Bit Commitment (BC) and Coin Tossing
(CT) protocols [4,5]. Quantum versions of these protocols were first
proposed by Bennett and Brassard [6].

BC is the information exchange protocol allowing two distant users
A and B which do not trust each other to implement the following scheme.
User A sends some (part of) information on his secret bit $b$ ($b=0$ or 1,
commitment stage) to user B in such a way that user B cannot recover
the secret bit chosen by A on the basis of information supplied alone.
However, this information should be sufficient to prevent cheating by user A,
i.e., later (at the disclosure stage) when user B asks user A to send him 
the rest information on the chosen secret bit, user A should be unable to 
change his mind and modify the value of his secret bit. The CT protocol is 
the scheme allowing two distant users which do not trust each other to 
implement the procedure of drawing an honest lot.

Classical versions of these protocols are based on unproved computational
complexity of some trap-door functions which require exponentially large
resources to calculate their inverse on the classical computer [7,8].

Some time ago it was generally assumed that the quantum protocols based on 
the fundamental restriction imposed by the laws of quantum mechanics rather 
than on the computational complexity are unconditionally secure [9]. However, 
it was later shown by Mayers, Lo and Chau [10,11] that the non-relativistic 
quantum BC protocol is not actually secure. User A can cheat user B without 
being detected by the latter employing the so-called EPR-attack (EPR stands 
for Einstein, Podolsky, and Rosen [12]). The possibility of successful 
EPR-attack is actually based on the result of paper by Hougston, Josza, and 
Wotters on the measurements performed over the quantum ensembles of composite 
systems [13].

All the above mentioned non-relativistic quantum protocols are only based
on the properties of the quantum states in the Hilbert space and do not 
explicitly contain the effects of state propagation between the two distant
users. However, actually the information transfer occurs in the Minkowski
space-time. Explicit accounting for this circumstance extends the
possibilities for development of new relativistic quantum protocols [14]
and substantially simplifies the proof of their security [15]. Restrictions
imposed by the special relativity on the measurements performed on quantum 
states allow to realize the secure BC and CT protocols in the ideal 
channel [16]. Failure of the EPR-attack in the relativistic case is related
to the impossibility of an instant modification of an extended quantum state.
In addition, it is even impossible to instantly and reliably distinguish 
between two orthogonal states. Restrictions imposed by the special 
relativity on the measurement of quantum states were first discussed
by Landau and Peierls [17].

In the present paper we propose the first relativistic BC protocol in
a quantum noisy channel. Intuitively, the idea behind the protocol is very 
simple. User A prepares (turns on the source) one of the two orthogonal 
states corresponding to 0 or 1 which are sent with the maximum possible 
speed (the speed of light $c$; further on we assume $c=1$) into the 
communication channel as they are being formed. As long as the states
are not fully accessible to user B, he cannot reliably determine the
value of the secret bit. User A cannot influence (again because of the
existence of the maximum propagation velocity) the part of the state
which has already left his laboratory and propagates through the communication 
channel (the commitment stage). When the state becomes fully accessible to
user B he can reliably determine the secret bit value (because of the 
orthogonality of the states) and compare it with that declared by user A
through the classical channel at the disclosure stage. Restrictions imposed
by the special relativity on quantum measurements allow to explicitly
realize the original idea of the Bit Commitment protocol on providing
only a part of information on the secret bit to the other party while 
spatial restriction of the state accessibility automatically results
in the restriction of accessible part of the Hilbert state space even
for ``internal'' degrees of freedom of the quantum system (e.g., spin or 
polarization) since they do not exist separately from the spatial 
degrees of freedom.

The protocol employs a pair of single-photon states with orthogonal
polarizations and the spatial amplitude of a special form corresponding 
to 0 and 1:
\begin{equation}
|\psi_{0,1}\rangle=\int_{0}^{\infty} dk{\cal F}(k)a^+(k)|0\rangle
\otimes |e_{0,1}\rangle=\int_{0}^{\infty} dk{\cal F}(k)|k\rangle
\otimes |e_{0,1}\rangle=|{\cal F}\rangle\otimes |e_{0,1}\rangle,
\end{equation}
where $a^+(k)$ is the creation operator for the state with momentum (energy) 
$k>0$, ${\cal F}(k)$ is the amplitude in $k$-representation, 
$|e_{0,1}\rangle$ is the polarization state, and
\begin{equation}
\int_{0}^{\infty} dk|{\cal F}(k)|^2=1,\quad [a(k),a^+(k')]=\delta(k-k'),\quad
\langle e_i|e_j\rangle=\delta_{ij},\quad i,j=0,1,\quad k\in(0.\infty).
\end{equation}
In the spatio-temporal $\tau$-representation the states are written as
\begin{equation}
|\psi_{0,1}\rangle=\int_{-\infty}^{\infty} d\tau{\cal F}(\tau)|\tau\rangle
\otimes |e_{0,1}\rangle,\quad
{\cal F}(\tau)= \int_{0}^{\infty} dk{\cal F}(k)\mbox{e}^{-ik\tau},
\langle k|\tau\rangle=\frac{\mbox{e}^{ik\tau}}{\sqrt{2\pi}},\quad
\tau=t-x,\quad \tau\in(-\infty,\infty),
\end{equation}
where ${\cal F}(\tau)$ is the amplitude in $\tau$-representation reflecting
the intuitive picture of a packet propagating in the positive direction
of $x$-axis with the speed of light and having the spatio-temporal shape
${\cal F}(\tau)$. The normalization condition in the $\tau$-representation
has the form [18]
\begin{equation}
\langle\psi_{0,1}|\psi_{0,1}\rangle=\langle{\cal F}|{\cal F}\rangle=
\int_{-\infty}^{\infty} \int_{-\infty}^{\infty} d\tau d\tau'
{\cal F}(\tau){\cal F}^*(\tau')[\frac{1}{2}\delta(\tau-\tau')+\frac{i}{\pi}
\frac{1}{\tau-\tau'}]= \int_{-\infty}^{\infty}|{\cal F}(\tau)|^2d\tau,
\end{equation}
\begin{displaymath}
\int_{-\infty}^{\infty}\mbox{e}^{ik\tau}\frac{1}{\tau+a}=
i\pi\mbox{ }\mbox{sgn}(k)\mbox{e}^{-ika}.
\end{displaymath}
Important for the proposed protocol are the following two circumstances:
1) There exists a maximum state propagation speed; 2) Orthogonal states
cannot be reliably distinguished when they are not fully accessible
(even if they remain orthogonal when restricted to the domain accessible 
to measurements). The classical bit values of 0 and 1 correspond to 
two orthogonal polarization states $|e_0\rangle$ and $|e_1\rangle$. 
Since the polarization does not exist separately from the spatial 
degrees of freedom ${\cal F}(\tau)$, the reliable (with probability 1)
distinguishability requires the access to the entire spatial domain
where the amplitude ${\cal F}(\tau)$ is different from zero. To be more 
precise, any measurement in a finite domain $\tau$ necessarily involves
a non-zero error probability in the state distinguishability. Generally, any 
measurement is described by an identity resolution in ${\cal H}$ [19--22,25], 
and when only a finite domain $\Delta(\tau)$ ($\overline{\Delta}(\tau)$ being
the complement to the entire space $\tau\in(-\infty,\infty)$) is accessible
to measurement the identity resolution has the form
\begin{equation}
I=\int_{-\infty}^{\infty}d\tau |\tau\rangle\langle \tau|\otimes I_{C^2}=
\int_{\Delta(\tau)} d\tau |\tau\rangle\langle \tau|\otimes
({\cal P}_0+{\cal P}_1)+
\int_{\overline{\Delta}(\tau)} d\tau |\tau\rangle\langle \tau|\otimes I_{C^2},
\quad {\cal P}_{0,1}=|e_{0,1}\rangle\langle e_{0,1}|,
\end{equation}
where ${\cal P}_{0,1}$ are the projectors to the polarization states
$|e_{0,1}\rangle$. If the measurement outcome occurs in the accessible domain
$\Delta(\tau)$, the probabilities of outcomes in the two orthogonal channels
${\cal P}_0$ and ${\cal P}_1$ are
\begin{equation}
\mbox{Tr}\{\rho(0,1)(I(\Delta(\tau))\otimes {\cal P}_{0,1})\}=
\int_{\Delta(\tau)} d\tau |{\cal F}(\tau)|^2=N(\Delta(\tau)),
\quad
\mbox{Tr}\{\rho(0,1)(I(\Delta(\tau))\otimes {\cal P}_{1,0})\}\equiv 0,
\end{equation}
where $\rho(0,1)=|\psi_{0,1}\rangle\langle\psi_{0,1}|$, and $N(\Delta(\tau))$
is the fraction of outcomes in the accessible domain. The probability of error
in that case is zero because of the orthogonality of the channels
$p_e(\Delta(\tau))=0$. However, if the outcome is not obtained
in the domain accessible to the measurement, the error probability is
$p_e(\overline{\Delta}(\tau))=1/2$, and the fraction of these outcomes is
\begin{equation}
\mbox{Tr}\{ \rho(0,1)\left(I(\overline{\Delta}(\tau))\otimes I_{C^2}\right) \}=
\int_{\overline{\Delta}(\tau) } d\tau |{\cal F}(\tau)|^2=
N(\overline{\Delta}(\tau)).
\end{equation}
The total error probability is
\begin{equation}
P_e=p_e(\Delta(\tau)) N(\Delta(\tau))+
p_e(\overline{\Delta} (\tau)) N(\overline{\Delta}(\tau))=
0\cdot N(\Delta(\tau))+ \frac{1}{2}\cdot N(\overline{\Delta}(\tau))=
\frac{1}{2}\int_{\overline{\Delta}(\tau)} d\tau |{\cal F}(\tau)|^2\neq 0.
\end{equation}
The protocol employs the states with a special spatio-temporal amplitude
corresponding to a state consisting of two strongly localized and separated
by an interval $\tau_0$ ``halves''
\begin{equation}
{\cal F}(\tau)=\frac{1}{\sqrt{2}}[f(\tau)+f(\tau-\tau_0)],\quad
\int_{-\Delta\tau}^{\Delta\tau}d\tau |f(\tau)|^2=
\int_{-\Delta\tau+\tau_0}^{\Delta\tau+\tau_0}d\tau |f(\tau-\tau_0)|^2=
1-\delta,\quad \delta\ll 1,\quad \Delta\tau\ll \tau_0,
\end{equation}
where $\delta$ can be chosen arbitrarily small. 
The amplitude $f(\tau)$ cannot possess a finite support [23], although
it can be arbitrarily strongly localized and can have a decay rate
arbitrarily close to the exponential one [23,24]. In the following we shall 
for brevity omit the parameter $\delta$ bearing in mind that it can be safely 
made the smallest parameter in the problem. The latter means that if the
accessible domain of the space-time $\tau$ covers the interval
$-\Delta\tau<\tau<\Delta\tau+\tau_0$, the error probability (8) is
$P_e=0$.  On the contrary, if only one half of the state is accessible,
the error probability (8) is $P_e=1/4$. In other words, this means
that reliable distinguishability of a pair of states (9) requires
access to the spatio-temporal domain of size $\approx \tau_0$ which,
because of the existence of the limiting propagation speed, 
cannot be achieved faster than $\tau_0$.

The input states sent by user A into the quantum communication channel
are $\rho_{in}(0,1)=(|e_{0,1}\rangle\otimes|{\cal F}\rangle)   
(\langle{\cal F}|\otimes\langle e_{0,1}|)$. Description of the quantum 
communication channel actually reduces to specifying the instrument
(sometimes also called ``superoperator'') [19--22,25] mapping the input
density matrices into the output ones (not necessarily normalized). Any 
quantum communication channel defines an affine mapping of the set of input 
density matrices into the set of output density matrices. Any mapping of that 
kind reduces to specifying the instrument $\mbox{\boldmath ${\cal T}$}$, 
\begin{equation}
\rho_{out}(0,1)=\mbox{\boldmath ${\cal T}$}[\rho_{in}(0,1)]=
\int_{-\infty}^{\infty}\int_{-\infty}^{\infty} d\tau d\tau'
\rho_{out}(\tau,\tau')|\tau\rangle\langle\tau'|\otimes \rho(e_0,e_1) =
\sum_{i=1}^{\infty}\lambda_i (|e_{i,0,1}\rangle\otimes |u_i\rangle)
(\langle u_i|\otimes\langle e_{i,0,1}|),
\end{equation}
where $|u_i\rangle=\int_{-\infty}^{\infty} d\tau u_i(\tau)|\tau\rangle$ are
the eigenvectors of the output density matrix operator (kernel
$\rho_{out}(\tau,\tau')$). Taking into account Eq. (4) one has
\begin{equation}
\int_{-\infty}^{\infty}  d\tau' \rho_{out}(\tau,\tau')u_i(\tau') =
\lambda_i u_i(\tau),\quad
\int_{-\infty}^{\infty} d\tau  u_i(\tau)u_j^*(\tau)=\delta_{ij},\quad
\sum_{i=1}^{\infty}\lambda_i\le 1.
\end{equation}
The output polarization vectors are
$|e_{i,0,1}\rangle=\alpha_{i,0,1}|e_0\rangle+\beta_{i,0,1}|e_1\rangle$
($|\alpha_{i,0,1}|^2+|\beta_{i,0,1}|^2=1$). Any instrument can be presented 
in the form $\mbox{\boldmath ${\cal T}$}[\rho]=\sum_{i}V_i\rho V_i^+$,
with $\sum_iV_iV_i^+\le I$ [19--22] (it is sufficient here to restrict 
ourselves to the discrete outcome space $i$). In our case this representation
can be written in the form
\begin{equation}
\mbox{\boldmath ${\cal T}$}[\ldots]=\sum_{i=1}^{\infty}\lambda_i
(|e_{i,0,1}\rangle\otimes |u_i\rangle)(\langle e_{0,1}|\otimes\langle{\cal F}|)
[\ldots]
(|{\cal F}\rangle\otimes |e_{0,1}\rangle)(\langle u_i|\otimes\langle e_{i,0,1}|)+
\mbox{\boldmath ${\cal T}$}_{\bot}[\ldots],
\end{equation}
where $\mbox{\boldmath ${\cal T}$}_{\bot}[\ldots]$ is the part of the 
instrument yielding identical zero on the subspace spanned by the vectors
$|{\cal F}\rangle\otimes|e_{0,1}\rangle$.

Writing the instrument in the form of Eq. (10) we assumed that the
decoherence of both basic polarization states occurs in the same way which 
is true if the medium does not possess gyrotropic properties. The latter 
condition is normally satisfied for optical fiber communication channels. 
However, if the decoherence of the states with different polarizations
occur in different ways and depends on the spatial degrees of freedom,
the following analysis can easily be extended to that case.

Since the time of preserving the secret bit is determined by the state 
extent ($\tau_0$) the channel length can be arbitrary; therefore,
we shall set it equal to zero without loss of generality. Actually,
the specification of the instrument is the description of the quantum
communication channel just as in the classical case where the probability 
distributions on the output alphabet is specified for each symbol of the 
input alphabet. At the intuitive level this mapping can be understood
(with some reservation) as the transformation of an input state
$|\psi_{0,1}\rangle$ with the shape ${\cal F}(\tau)$
and polarization $e_{0,1}$ into one of the output states with the
shape $u_i(\tau)$ and polarization $e_{i,0,1}$ occurring with the
probability $\lambda_i$. The fact that the sum of probabilities
does not exceed unit, $\sum_i\lambda_i\le1$, can be interpreted in our case
as the disappearance (absorption) of a photon in the channel. The channel
properties are determined by the functions $u_i(\tau)$ and probabilities 
$\lambda_i$ which are assumed to be known from the {\it a priori}
considerations (and can be found from the channel calibration procedure).
If it is possible to choose a new interval of the state halves localization
at the output $D\tau$ such that
\begin{equation}
\forall\mbox{ }i=1,\infty,\quad
\frac{1}{2}\int_{-D\tau}^{D\tau}d\tau |u_i(\tau)|^2=
\frac{1}{2}-\delta,\quad
\frac{1}{2}\int_{-D\tau+\tau_0}^{D\tau+\tau_0} d\tau |u_i(\tau)|^2=
\frac{1}{2}-\delta,\quad D\tau\ll\tau_0,
\end{equation}
where $\delta$, just as previously (9), is arbitrarily small, the channel is 
suitable to the realization of the proposed protocol. In other words,
the channel has the property that the strongly localized states at the input
still remain strongly localized at the output to within $D\tau\ll\tau_0$ and 
$D\tau>\Delta\tau$ (fig.1), although they can change their shape and 
polarization. The quantity $D\tau$ then determines the accuracy with which
user B can detect the delay of choice of secret bit by user A (delay of
sending the state in the communication channel). The probability of
detecting a state at the output by user B in the spatio-temporal window
$\Delta(\tau)$ covering only one of the halves $u_{i}(\tau)$ 
independently of the outcome in the channels ${\cal P}_{0,1}$ is
\begin{equation}
\mbox{Pr}\{\Delta(\tau)\}=\mbox{Tr}\{
\mbox{\boldmath ${\cal T}$}[\rho_{in}(0,1)] \left(I(\Delta(\tau)\otimes
I_{C^2} \right)\}=\sum_{i=1}^{\infty}\lambda_i\int_{\Delta(\tau)}d\tau
|u_{i}(\tau)|^2\le (\frac{1}{2}-\delta)\sum_{i=1}^{\infty} \lambda_i\le
\frac{1}{2}-\delta\le\frac{1}{2},
\end{equation}
and can be made arbitrarily close (with the exponential accuracy by suitably
choosing $D\tau$ and $\tau_0$) to 1/2. In this case, the probability of
correct identification of the state when only one half of the state is 
accessible (i.e. during time interval $\approx\tau_0$)
does not exceed $1/2\cdot 1/2=1/4$ (8).

Now we shall calculate the probability of error for the case when the states
become fully accessible (after the time $D\tau+\tau_0\approx\tau_0$ elapses; 
for the ideal communication channel the distinguishability error is zero).
If the state is fully accessible (after time $\approx \tau_0$ since the 
protocol was started) the probability of an outcome in one of the 
channels ${\cal P}_{0,1}$ is
\begin{equation}
\mbox{Pr}\{\Delta(\tau)+\overline{\Delta}(\tau)\}=\mbox{Tr}\{
\mbox{\boldmath ${\cal T}$}[\rho_{in}(0,1)] \left(I(\Delta(\tau)\otimes
I_{C^2} \right)\}=\sum_{i=1}^{\infty}\lambda_i\le 1.
\end{equation}
The fact that $\mbox{Pr}\{\Delta(\tau)+\overline{\Delta}(\tau)\}\le 1$
means that not all states reach the channel output, i.e. the states are absorbed 
in the channel with the probability $1-\sum_{i=1}^{\infty}\lambda_i$ 
(formally, this is the probability for a state to never become accessible
for user B). In that case, where the measuring apparatus employed by user B
did not fire at all, he can only guess which state was actually sent, the 
contribution to the error probability from these events being
$1/2(1-\sum_{i=1}^{\infty}\lambda_i)$. Let us now calculate the contribution
to the error probability from the events when the measuring apparatus
employed by user B produced some outcome. The measurement minimizing
the polarization distinguishability error for the two ``honest'' input 
states sent by A is given by the following identity resolution
(for detail, see e.g. Ref.[26]):
\begin{equation}
\sum_{i=1}^{\infty}{\cal P}_i\otimes (E_0+E_1)+{\cal P}_{\bot}\otimes I_{C^2}=
I\otimes I_{C^2},\quad {\cal P}_i=|u_i\rangle\langle u_i|,\quad
{\cal P}_{\bot}=I-\sum_{i=1}^{\infty} {\cal P}_i,
\end{equation}
\begin{equation}
E_0+E_1=I_{C^2}, \quad E_0=|\tilde{e}_0\rangle\langle\tilde{e}_0|,\quad
I_{C^2}=|e_0\rangle\langle e_0|+|e_1\rangle\langle e_1|,
\end{equation}
where $|\tilde{e}_0\rangle$ are the eigenvectors of the operator
\begin{equation}
\Gamma=\gamma_{00}|e_0\rangle\langle e_0|+\gamma_{01}|e_0\rangle\langle e_1|+
\gamma_{10}|e_1\rangle\langle e_0|+\gamma_{11}|e_1\rangle\langle e_1|,
\end{equation}
\begin{displaymath}
\gamma_{00}=\frac{1}{2}\sum_{i=1}^{\infty}\lambda_i
(|\alpha_{i,1}|^2-|\alpha_{i,0}|^2), \quad
\gamma_{11}=\frac{1}{2}\sum_{i=1}^{\infty}\lambda_i
(|\beta_{i,1}|^2-|\beta_{i,0}|^2),\quad
\end{displaymath}
\begin{equation}
\gamma_{01}=\frac{1}{2}\sum_{i=1}^{\infty}\lambda_i
(\alpha_{i,1}\beta_{i,0}^{*}-\alpha_{i,0}\beta_{i,1}^{*}),\quad
\gamma_{10}=\gamma_{01}^{*}.
\end{equation}
Taking into account Eq. (18) and bearing in mind that the states 0 and 1 are 
chosen by user A with equal {\it a priori} probabilities of 1/2, the total 
error for distinguishing between the polarizations of the two ``honest'' input 
states when they are fully accessible can be represented as 
\begin{equation}
P_e=\frac{1}{2}(1-\sum_{i=1}^{\infty}\lambda_i)+
\frac{1}{2}\mbox{Tr}\{\mbox{\boldmath ${\cal T}$}[\rho_{in}(0)]
((\sum_{i=1}^{\infty} {\cal P}_i)\otimes E_1)  \}+
\frac{1}{2}\mbox{Tr}\{\mbox{\boldmath ${\cal T}$}[\rho_{in}(1)]
((\sum_{i=1}^{\infty} {\cal P}_i)\otimes E_0)  \}=
\frac{1}{2}-|\gamma_2|<\frac{1}{2},
\end{equation}
where $\gamma_2$ is the negative eigenvalue of the operator $\Gamma$ in Eq. (18),
\begin{equation}
\gamma_2=\frac{1}{2}(\gamma_{00}+\gamma_{11})-\frac{1}{2}
\sqrt{(\gamma_{00}-\gamma_{11})^2+4|\gamma_{01}|^2 }.
\end{equation}
If the polarizations $|e_0\rangle$ are $|e_1\rangle$ disturbed in the channel 
in the same way, one has $\gamma_2=-|\gamma_{01}|$. For the ideal channel
Eqs. (18--20) yield $P_e=0$.

The protocol consist of the following steps.
1) The users control only their local neighbourhoods. They agree in advance on
the time when the protocol is started, the states (${\cal F}(\tau)$) employed,
and the adopted polarization basis $|e_{0,1}\rangle$) for 0 and 1.
2) User A encodes the secret bit $b$ (0 or 1) as the parity bit of $N$
states $\tilde{0}$ and $\tilde{1}$ consisting of the blocks each containing
$k$ bits ($b=\sum_{j=1}^{N}\oplus a[i,j]$,
$i=1..k$; all $a[i,j]$ belonging to the same block are identical) and
sends $k\cdot N$ states randomly distributed among $k\cdot N$ quantum 
communication channels. User B performs measurements described by Eq. (16).
3) At the disclosure stage, at any time $-\Delta\tau<\tau<\Delta\tau+\tau_0$,
user B can ask user A to announce through a classical communication channel
what he actually sent to user B. 4) After the protocol duration time elapses,
user B compares the outcomes of his measurements with the data obtained from
user A through the classical communication channel. 5) If all the tests
are successful, the protocol is completed; otherwise it is aborted.

Before the protocol duration time elapses completely, the probability
of correct secret bit identification by user B exceeds 1/2 (i.e. the 
probability of simple guessing) by only an exponentially small amount.
Indeed, if the block representation of 0 and 1 is adopted, the number 
of binary strings of length $k\cdot N$ is (see Ref.[27] for the details 
of summation)
\begin{equation}
N_{odd}=N_{even}=\frac{1}{2}\sum_{m=0}^{N} C^{m\cdot k}_{N\cdot k}=
\frac{2^{N\cdot k}}{2k}\sum_{l=1}^{k} \cos^{N\cdot k}{(\frac{l\pi}{k})}
\cos{(lN\pi)} \approx\frac{1}{2k} 2^{N\cdot k},
\end{equation}
which practically coincides with the total number of binary strings of length
$N\cdot k$. The Shannonn information [28--30] of the set of block strings is
(to within the rounding) the number of binary digits required to identify the 
string parity,
\begin{equation}
I=\mbox{log}_2\left(\frac{2^{N\cdot k}}{2k}\sum_{l=1}^{k}
\cos^{N\cdot k}{(\frac{l\pi}{k})}\cos{(lN\pi)} \right)\approx \eta \mbox{ }
N\cdot k ,\quad \eta \approx 1,
\end{equation}
i.e. one should know almost all bits in the string. However, if only one half 
of the state is accessible ($\Delta\tau<\tau<\Delta\tau+\tau_0$), the
error probability for determination of any particular bit in the string is
not less than 1/4 even in a noiseless channel (see Eq. (8)). Therefore, the 
probability for user B to learn the parity bit before the protocol duration 
time completely elapses does not exceed
\begin{equation}
P(parity)=\frac{1}{2} + 2^{-\frac{\eta}{2}N\cdot k}.
\end{equation}
We shall now calculate the probability of correct identification of the
parity bit after the protocol duration time fully elapsed. The block 
representation with $k$ bits is stable (the errors are corrected by
majority voting), if the number of errors in each block does not exceed
$k/2-1$. The probability of wrong identification of a block-wise
$\tilde{0}$ or $\tilde{1}$ is
\begin{equation}
P_e(k)=\sum_{i=k/2}^{k} C^i_k P_e^i (1-P_e)^{k-i}\approx
\sqrt{\frac{2}{\pi k}} [2\sqrt{P_e(1-P_e)}]^k,
\end{equation}
which can be made arbitrarily small by appropriate choice of $k$.
The total error in the parity bit identification is (we assume $N$ to be even)
\begin{equation}
P_e(parity)=\sum_{i=odd}^{N-1} C^i_N P_e^i(k) (1-P_e(k))^{N-i},
\end{equation}
where summation is performed over the odd subscripts $i$ only since
the error in the calculated parity bit arises when an odd number of blocks
are wrongly identified.
Making use of
\begin{equation}
\frac{1}{2}[(x+y)^N-(x-y)^N]=\sum_{i=odd}^{N-1} C^i_N x^i y^{N-i},
\end{equation}
and substituting $x=P_e(k)$ and $y=1-P_e(k)$ ($x+y=1$), one obtains
\begin{equation}
P_e(parity)=\frac{1}{2}[1-(1-2P_e(k))^N].
\end{equation}
By appropriate choice of $k$, for a specified quantum communication channel 
the probability $P_e(k)$ can be made arbitrarily small such that the 
quantity $NP_e(k)\ll 1$ is exponentially small. Under these conditions
the probability of wrong parity bit identification after the protocol 
duration time elapses is also arbitrarily small so that the probability of 
correct is arbitrarily close to 1.

Let us now discuss the protocol stability against cheating by user A.
Since the minimal Hemming distance between the two block-wise strings
with different parities is $k$ (minimal number of non-coinciding bits),
alteration of the string parity requires modification of at least $k$ bits.
Since the probability of correct identification of each block-wise
$\tilde{0}$ or $\tilde{1}$ is not worse than 
$1-P_e(k)\rightarrow 1$ (see Eq. (25), $P_e(k)$ is exponentially small), 
the probability of undetectable cheating by user A does not exceed this 
quantity.

The protocol is also stable against the delay of choice of secret bit
by user A. Note that for the ``honest'' non-delayed input states
the probability of the outcome in the channel $\bot$,
${\cal P}_{\bot}=I-\sum_{i=1}^{\infty}{\cal P}_i$ is zero:
\begin{equation}
\mbox{Pr}(\Delta(\tau)+\overline{\Delta}(\tau))=
\mbox{Tr}\left\{ (\mbox{\boldmath ${\cal T}$}[\rho_{in}(0,1)]+
\mbox{\boldmath ${\cal T}$}_{\bot}[\rho_{in}(0,1)])
({\cal P}_{\bot}\otimes I_{C^2})\right\} =
\end{equation}
\begin{displaymath}
\mbox{Tr}\left\{
\left(
\left( \sum_i \lambda_i|\alpha_{i,0,1}|^2{\cal P}_i\right)
\otimes |e_0\rangle\langle e_0| +
\left( \sum_i \lambda_i \alpha_{i,0,1}\beta_{i,0,1}^{*} {\cal P}_i\right)
\otimes |e_0\rangle\langle e_1| +
\right.\right.
\end{displaymath}
\begin{displaymath}
\left.\left.
\left( \sum_i \lambda_i \beta_{i,0,1}\alpha_{i,0,1}^{*} {\cal P}_i\right)
\otimes |e_1\rangle\langle e_0| +
\left( \sum_i \lambda_i|\beta_{i,0,1}|^2{\cal P}_i\right)
\otimes |e_1\rangle\langle e_1|
\right) (I-\sum_j {\cal P}_j)\otimes I_{C^2}
\right\} =0,
\end{displaymath}
since $|\alpha_{i,0,1}|^2+|\beta_{i,0,1}|^2=1$ and
${\cal P}_i{\cal P}_j=\delta_{ij}{\cal P}_i$.

Any delay of the input state for more than $D\tau$ can be detected
with the probability arbitrarily close for 1. To prove this statement, 
we shall need the requirements imposed on the instrument (12) by the special
relativity (to be more precise, by the existence of the maximum propagation 
speed). If a strongly localized state (in the sense that its amplitude 
$\mu(\tau)$ satisfies the equation
$\int_{-\Delta\tau}^{\Delta\tau}d\tau |\mu(\tau)|^2=1-\delta$, $\delta$ 
being exponentially small, and $\Delta\tau\rightarrow 0$) is prepared at the 
input of an arbitrary quantum communication channel, then this state cannot
be detected at the output of the channel in time less than $t=L/c$
(to be more precise, the detection will take place within the time interval
$-\Delta\tau+L/c\le t\le \Delta\tau+L/c$ 
with probability arbitrarily close to 1, where $L$ is the channel length).
In our case the instrument (12) should map the states prepared at the channel 
input at later times into the states that arise at the output also at later 
times. The delay of the state amplitude leading front at the output
cannot be less than its delay at the input.

Any delayed input state can be written as (we omit the polarization degrees
of freedom for brevity)
\begin{equation}
\rho_{delay}=\sum_{l}\mu_l
|\mu_l\rangle\langle\mu_l|,\quad \sum_l \mu_l=1,\quad
|\mu_l\rangle=\int_{-\infty}^{\infty}d\tau \mu_l(\tau)|\tau\rangle,
\end{equation}
where $|\mu_l\rangle$ are the density matrix eigenvectors and the supports of 
functions $\mu_l(\tau)$ do not overlap in the interval $D\tau$ with the
support of the leading halves of the functions $u_i(\tau)$ arising at the 
channel output from the non-delayed states. At the channel output
$\rho_{delay}$ will be transformed into the density matrix whose eigenstates
$|\eta_k\rangle$ have the supports which also do not overlap with the
front half of $u_i(\tau)$ in the interval $D\tau$:
\begin{equation}
(\mbox{\boldmath ${\cal T}$} +
\mbox{\boldmath ${\cal T}$}_{\bot})[\rho_{delay}]=
\sum_k \eta_k |\eta_k\rangle\langle \eta_k|,
\quad \sum_k \eta_k\le 1,\quad |\eta_k\rangle =
\int_{-\infty}^{\infty} d\tau \eta_k(\tau)|\tau\rangle.
\end{equation}
This implies that $|\langle \eta_k|u_i\rangle|^2\le1/2$ since 
$\eta_k(\tau)$ does not cover the front half of $u_i(\tau)$ where
half of the norm (i.e., 1/2) of the state $u_i(\tau)$ is localized.

For the delayed states the probability of the outcome in the channel
$\sum_i {\cal P}_i<I$ is
\begin{equation}
\mbox{Tr}\left\{
\left(\sum_k \eta_k |\eta_k\rangle\langle \eta_k|\right)
\left(\sum_i{\cal P}_i\right) \right\} < 1,
\end{equation}
while for the non-delayed states this probability is 1. Similarly, the
probability of the outcome in the channel 
${\cal P}_{\bot}=I-\sum_{i=1}^{\infty}{\cal P}_i$ (note that for
``honest'' states this probability is zero) is
\begin{equation}
\mbox{Tr} \left\{
\left(\sum_k \eta_k |\eta_k\rangle\langle \eta_k|\right)
\left(I-\sum_i{\cal P}_i\right) \right\}=p_{\bot}\neq 0.
\end{equation}
The sum of probabilities for both channels is 1 if all states
reach the channel output (are not absorbed, i.e. $\sum_k \eta_k=1$).

Possible delay in the choice of the secret bit (delay of the state) 
is detected through the appearance of outcomes in the channel 
${\cal P}_{\bot}$ with the probability $p_{\bot}$. To change the parity bit, 
it is sufficient to delay the states in only one block containing $k$ bits.
The probability for user A to delay $k$ states and remain undetected is equal 
to the probability of an event when all $k$ delayed states do not give
a single outcome in the channel ${\cal P}_{\bot}$ thus imitating the 
measurements statistics for ``honest'' states. We have
\begin{equation}
P_{cheat}=(1-p_{\bot})^k\ll 1,
\end{equation}
which can be achieved for any specified $p_{\bot}$ by choosing a sufficiently 
large $k$.

Thus, the protocol allows to realize the honest bit commitment protocol
with the probability arbitrarily close to 1.

This work was supported by the Russian Fund for Basic Research 
(grant N 99-02-18127),  the project  ``Physical foundations of quantum
computer'' and the program ``Advanced technologies and devices of micro- 
and nanoelectronics'' (project N 02.04.5.2.40.T.50).

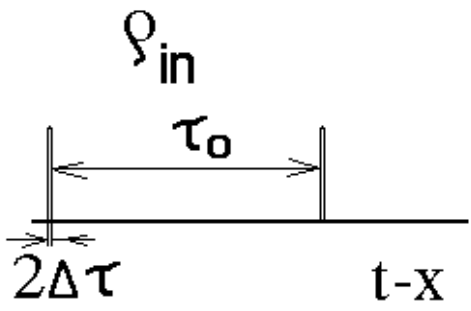
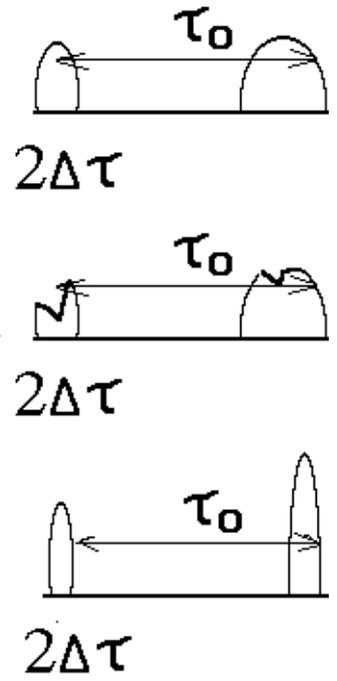
\end{document}